\documentclass[12pt]{iopart}
\usepackage{epsfig}

\newcommand{\pbar}{\overline{\rm p}}
\usepackage{iopams}  
\begin{document}

\title[An Experimental Overview of Results Presented at SQM 2006]{
An Experimental Overview of Results Presented at SQM 2006}

\author{R. K. Seto}

\address{Department of Physics, University of California, Riverside
, Riverside, CA, 92521, USA}
\ead{richard.seto@ucr.edu}
\begin{abstract}
I will present an overview of what was learned from the experimental results
presented at Strange Quark Matter 2006 concentrating primarily on RHIC data.
\end{abstract}


\section{Introduction}
I have been asked to give an critical overview on the experimental results 
shown in the conference with a emphasis of what has been learned
and the challenges that are ahead in trying to understand the physics of 
the strongly interacting quark-gluon plasma. I will not try to summarize
all of the  results presented, 
rather I will concentrate primarily on RHIC data from this conference. 
Throughout this summary, I will periodically review some of the
previous results for those not familiar with the present state of the field
{\cite{whitepapers}}.

As we came into this conference we knew several things:
\begin{itemize}
\item The system created in central heavy ion collisions at RHIC
is thermally and 
chemically equilibrated with thermalization times less than 2 fm/c.
\item The energy density at thermalization is very high, greater than 
10 GeV/fm$^3$.
\item The energy loss of fast partons is large.
\item The viscosity/entropy is very low, perhaps approaching that of a 
perfect fluid{\cite{stringtheory}}.
\end{itemize}

This basic picture has been stable for the past year. 
I will refer to the created system as a sQGP (Strongly Interacting Quark
Gluon Plasma). Much of the data presented 
in this conference has reinforced these basic ideas.
In addition some important theoretical insights have been gained, some of 
which will be mentioned in this review. 

RHIC is unique in its ability to accelerate a variety of nuclear
species over a large range of energies - $\sim$20 GeV center of mass to 200
GeV center of mass.  The dedicated nature of the machine together
with its ensemble of 4 experiments has given an unprecedented variety
of data sets. New results from one experiment, can quickly be confirmed 
and checked
by others. The capabilities of the 4 detectors are now being well
utilized giving results on an extraordinary variety of particle types
- with multi-strange baryons and charmed particles coming into the
mainstream.

\section{The evolution of the sQGP}

\begin{figure} 
  \begin{center}
    \leavevmode
    \epsfig{file=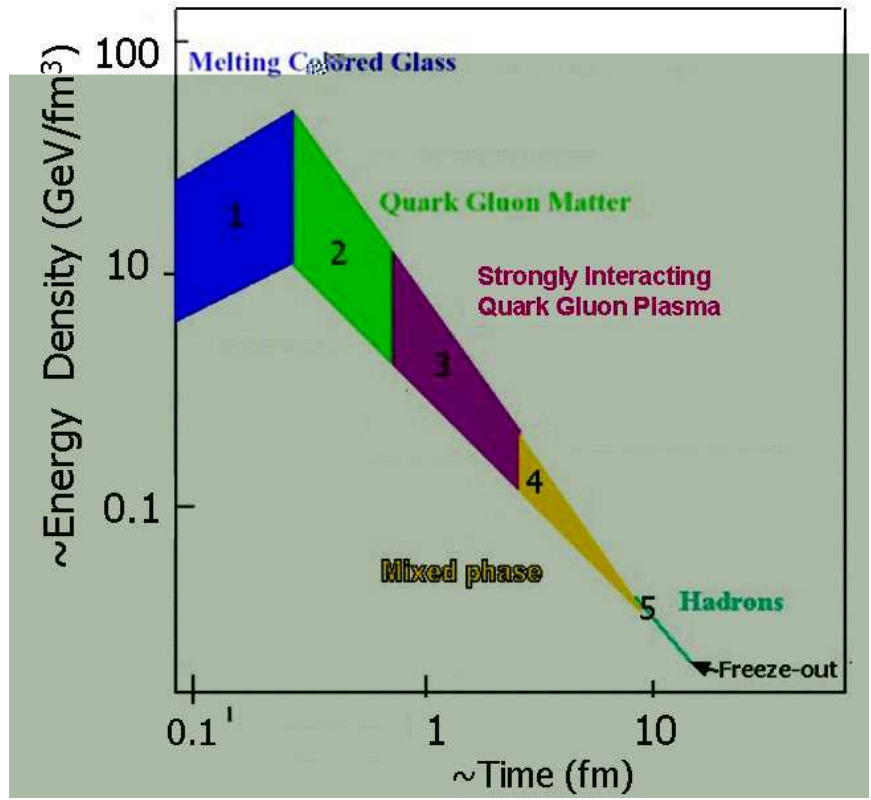, width=4.5in}
  \end{center}
  \caption{Time evolution of an AuAu collision at RHIC. The regions 1-5 are
explained in the text.}
  \label{fig:qgphistory}
\end{figure}

We are studying a dynamically evolving bulk system,
and experimental signatures give information on different times
of the collision history. To set the context, Fig.~\ref{fig:qgphistory} shows
 a cartoon of the evolution of the system with various stages
as follows:
\begin{description}
\item[Region 1:] the earliest moments of the collision where high momentum 
probes are formed. This early phase of the collision can be described by 
saturated parton functions - the Colored Glass Condensate. 
\item[Region 2:] thermalization is reached in region 2 at $\sim$0.6 fm.
\item[Region 3:] the era of the sQGP where much of the elliptic 
flow developers. Fast partons loose most of their energy during this portion 
of the collision.
\item[Region 4:] the mixed phase region. Recent lattice calculations 
indicate that the phase transition very likely a cross over, however, 
the temperature range over which the
transition occurs is narrow, $\sim$20 MeV, hence the for experimental concerns,
the transition can be assumed to be a first order transition. It is
this era where hadronization takes place - presumably via recombination. 
\item[Region 5:] the era of hadrons, where rescattering 
and freezeout occur. 
\end{description}
It is important to note that we have assumed a partonic phase for 
regions 1-4. All reasonable explanations of the data demand this. As we shall
see, further data continues to bear this out.

\section{Energy Density: Fast Parton Energy Loss}
\begin{figure}
\centering
\begin{tabular}{cc}
\epsfig{file=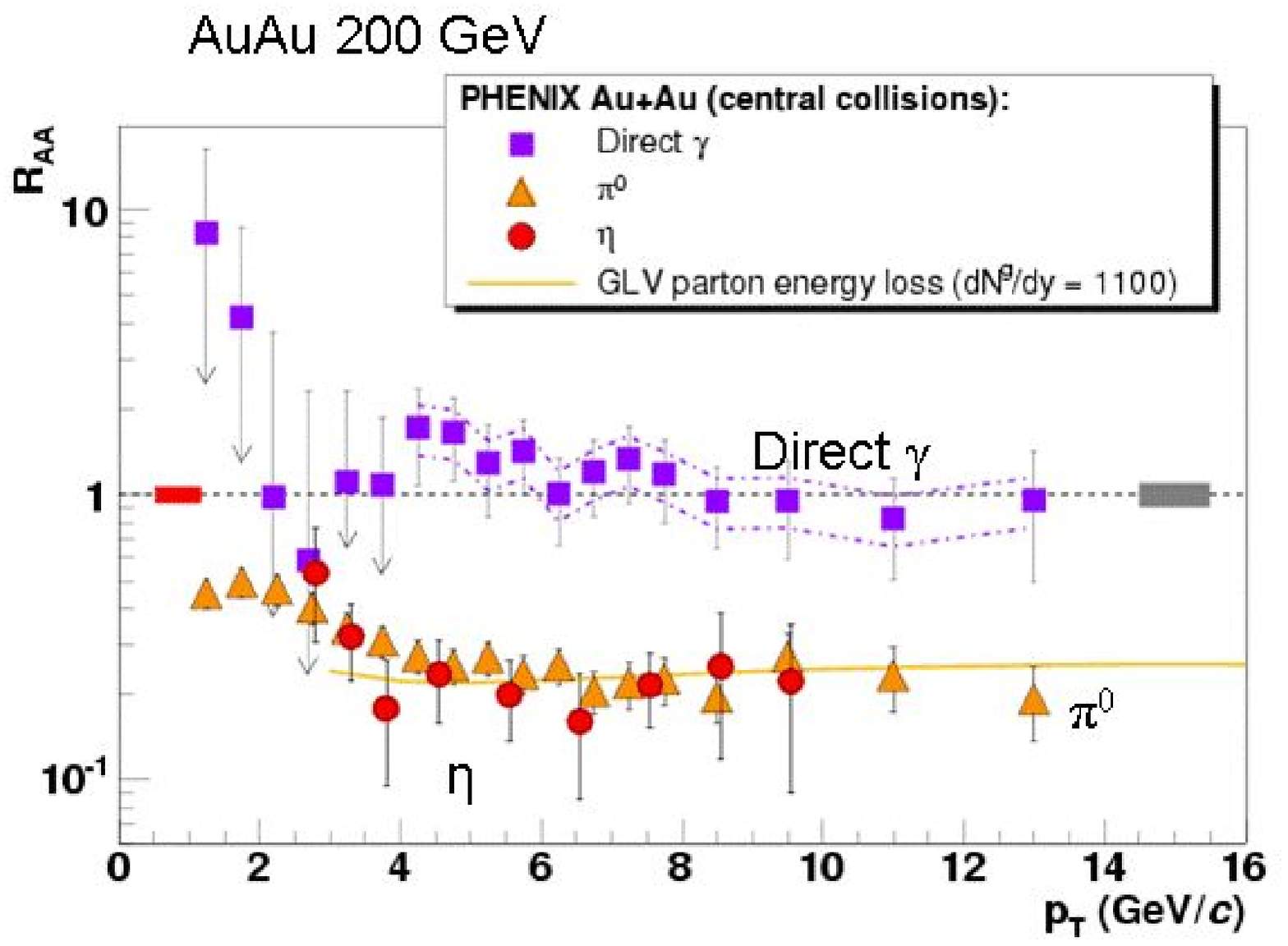,width=3.3in,clip=} & 
\epsfig{file=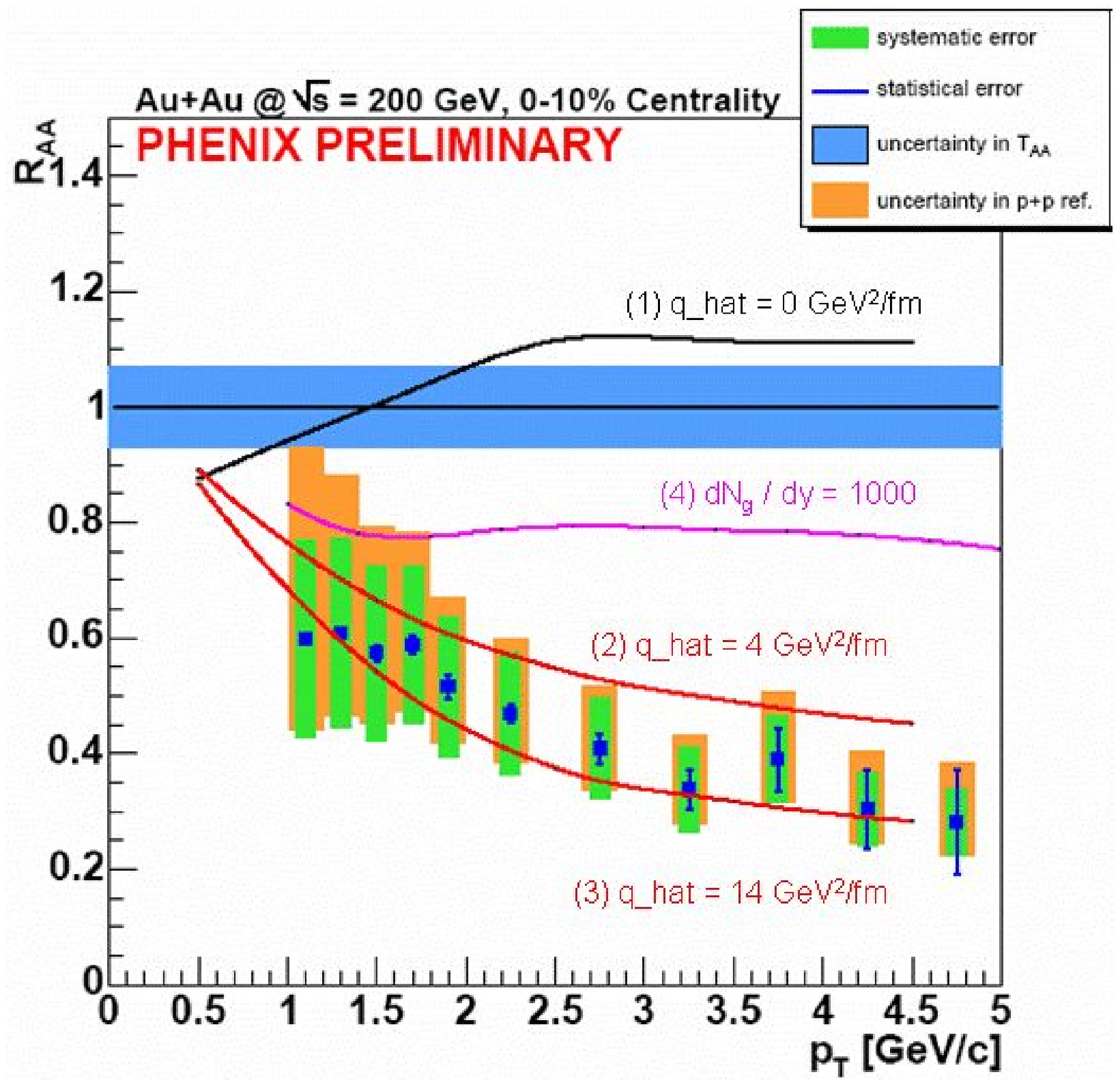,width=2.7in,clip=} \\
(a) & (b)\\
\end{tabular}
  \caption{(a) R$_{AA}$ of direct photons, $\pi^0$'s and $\eta$'s for central
AuAu collisions at $\sqrt{s}$=200 GeV as measured by PHENIX. (b) R$_{AA}$ for
non-photonic electrons. The lines correspond to models of energy loss: (1-3)
 are from N. Armesto et al, PRD 71 054027, and (4) is from M. Djordjevc, M. Gyulassy \& S. Wicks, PRL 94 112301.}
  \label{fig:jetquench}
\end{figure}

The phenomena of parton energy
loss or ``jet quenching'' is perhaps the most well known result from
RHIC. High momentum hadrons with p$_T \gtrsim$ 4
GeV/c are suppressed by a factor of 5 in central Au+Au
collisions, relative to pp collisions when scaled by the number
of binary collisions. Fast partons, the source of high
momentum hadrons, are produced very early in the collision (region 1 in
Fig.~\ref{fig:qgphistory}), and penetrate through the created matter
and fragment into hadrons. Quarks and gluons loose energy copiously
in the medium because of their color charge (primarily in region 3).  To
quantify the energy loss, we define
$$R_{AA}=\frac{1}{n_{coll}}\frac{d^2N_{AA}/dydp_T}{d^2N_{pp}/dydp_T}$$ 
where $n_{coll}$ is the
number of binary collisions calculated from a Glauber model in AuAu
collisions; $d^2N_{AA}/dydp_T$ and $d^2N_{pp}/dydp_T$ are the
differential yields observed in AuAu and pp collisions respectively.
This ratio should be unity if binary scaling holds.
Fig.~\ref{fig:jetquench}(a) shows the ratio R$_{AA}$ as measured by
the PHENIX collaboration for direct photons, $\pi^0$'s and $\eta$'s in
central Au+Au collisions. Direct photons show a ratio consistent with
unity as expected since they  suffer only a negligible amount of
energy loss.  However both $\pi0's$ and now $\eta's$ are shown to be
suppressed by a factor of 4-5. Calculations done assuming loss by
gluon radiation give gluon densities of $dN_{gluon}/dy$$\sim$1000 or energy
densities of 10-15 GeV/fm$^3$ at the time of initial
thermalization{\cite{wicks}}.  We will see that this is consistent with other
methods of calculating the initial energy density. This should be
compared with the expected critical energy density from lattice gauge
calculations of about 0.6 GeV/fm$^3$ {\cite{lattice}}.  It is
important to note here, that the energy loss calculations do not
require a deconfined medium, but depend only on the energy
density.

Dokshitzer and Kharzeev have shown that for heavy quarks, 
such as charm or bottom,
gluon radiation in a conical region around the momentum vector of the quark
would be suppressed thus limiting the energy loss of heavy quarks{\cite{deadcone}}. The effect could be rather significant;
a factor of $\sim$2 depending on the energy of the quark. Both PHENIX and STAR
have looked at so called ``non-photonic'' electrons, i.e. electrons
primarily from the prompt decay of heavy quarks. They have observed
a unexpectedly large suppression of such electrons 
Fig.~\ref{fig:jetquench}(b) {\cite{stjetquench}}. This has
led to a discussion of whether gluon
radiative energy loss is responsible for the entire suppression of high
momentum hadrons, or whether there may be additional mechanisms such
as collisional energy loss{\cite{wicks}\cite{energylosstheory}}.
However, as can be seen in the figure, a reasonable fit can be made
to the data using the so called BDMPS theory with
values of $\hat{q}$ between 4 and 14 GeV$^2$/fm. The higher value of 14
GeV$^2$/fm corresponds an
unrealistically high energy density. 
Since these measurements have been made using electrons, both c and b mesons
contribute; the energy loss for b quarks is considerably less than for
c quarks and it will be critical for experiments to examine the suppression
patterns separately for charm and bottom hadrons, in order to make a accurate
measurement of the energy loss and thus the energy density.

\section{Hadronization and the Idea of Recombination}

\begin{figure}
\centering
\begin{tabular}{cc}
\epsfig{file=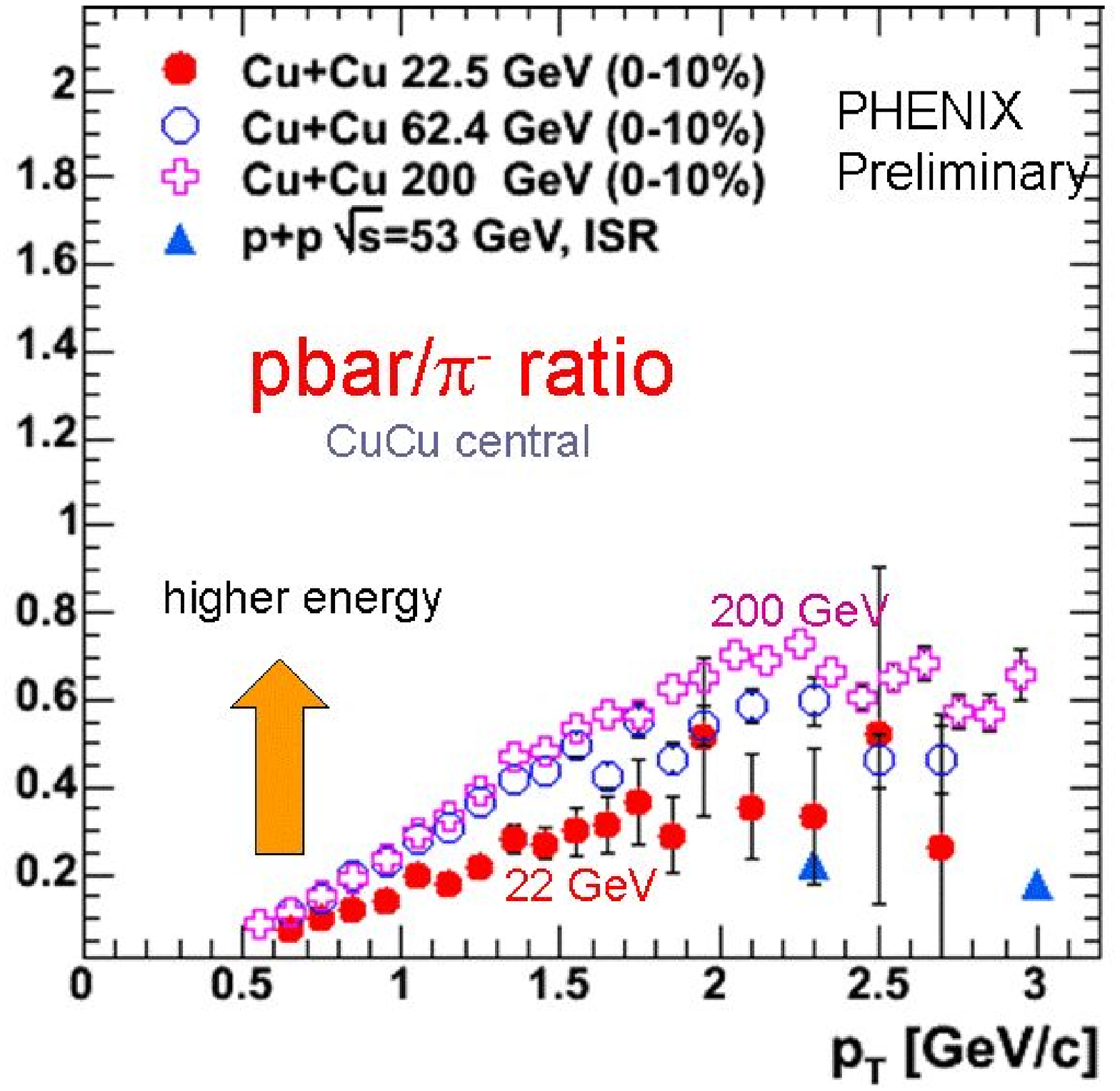,width=2.7in,clip=} & 
\epsfig{file=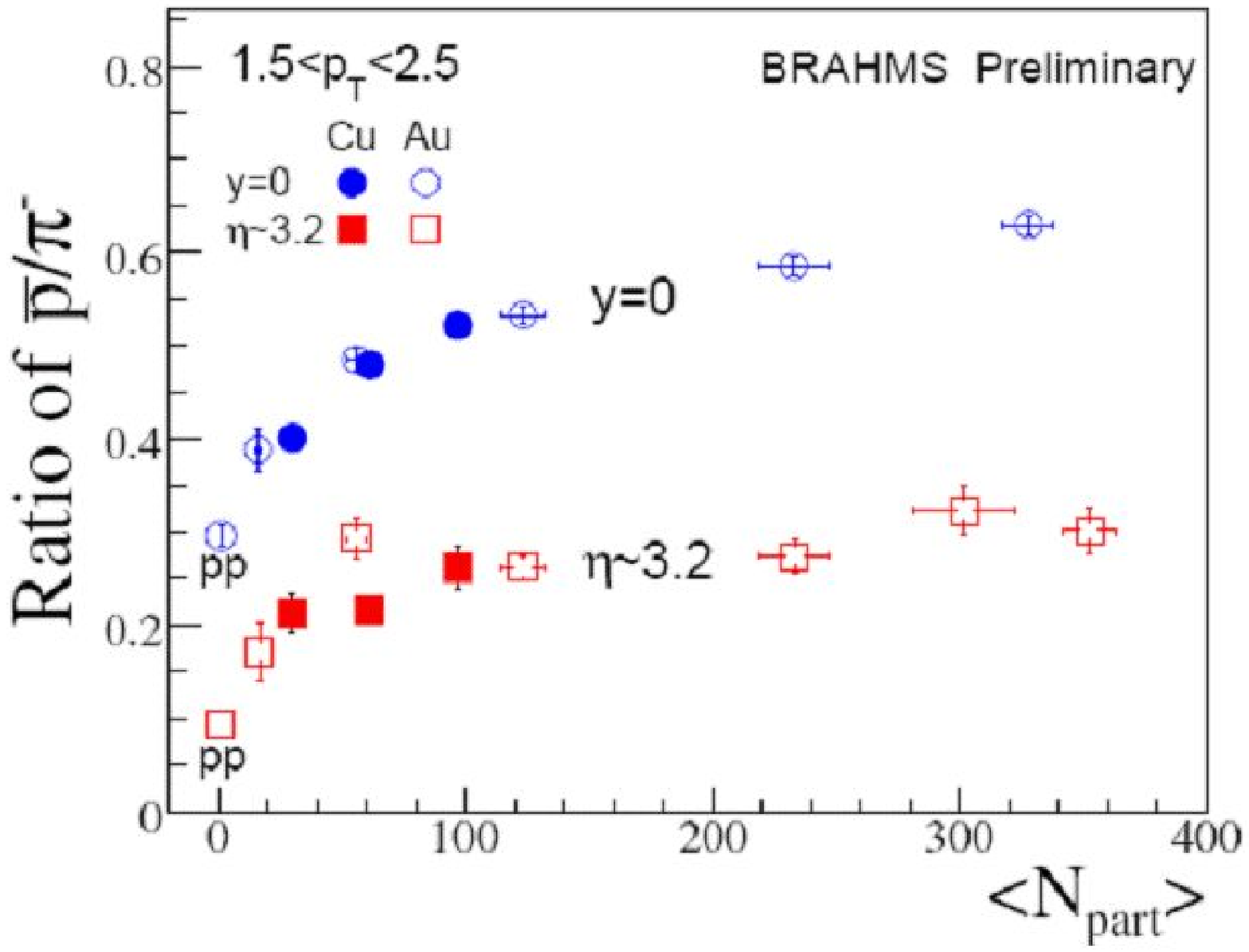,width=3.3in,clip=} \\
(a) & (b)\\
\end{tabular}
  \caption{(a)~$\pbar/\pi^-$ ratio in central CuCu collisions for a variety
of center of mass energies as a function of p$_T$. (b) $\pbar/\pi^-$ as a 
function of the number of participants, N$_{part}$ for AuAu and CuCu collisions at $\sqrt{s}$=200 GeV at mid and forward rapidities. }
  \label{fig:ptopi}
\end{figure}

One of the major surprises from RHIC was seen in data from identified
particles.  The ratio of baryons to mesons at moderate values of p$_T$
was much larger that in pp collisions, where high momentum hadrons
come from the fragmentation of quarks and gluons.  The effect was
largest in the most central AuAu collisions with $\pbar$ to $\pi^-$
$\sim$ 0.7, and was reduced for more peripheral collisions.  For
comparison, the $\pbar$ to $\pi^-$ ratio at the ISR was about 0.2{\cite{isr}}.
Additional data shown at this conference, confirms the effect in Cu+Cu
collisions. In making a direct comparison with the Au data,
(Fig.~\ref{fig:ptopi} (b)) we can see that the effect appears to
scale with N$_{part}$, the number of participating nucleons in the
collision. It is interesting to note that many different phenomena
seem depend on the colliding species only via N$_{part}$.  In
addition, the data shows that this baryon enhancement increases with
centrality, and decreases at forward rapidities.

Recombination as a mechanism for hadronization seems to explain this
result nicely{\cite{recomb}}. The idea is as follows. One assumes a
thermalized partonic system where the constituents which carry the
momentum distribution are of similar mass. Hadrons are formed when
these partons combine, nominally two primary constituents for mesons
and three for baryons. The average p$_T$ of mesons would then be
$\sim$2T, and for baryons, $\sim$3T where T is the temperature
of the partonic system.  This results in a spectrum of
baryons which is harder than for mesons, hence the baryon to meson
yield at a given p$_T$ can be considerably larger than in pp
collisions, where the primary mechanism of hadron production is the
fragmentation of partons.  The tendency to produce baryons rather than
mesons is related to the density of partons in the system with higher density
favoring baryons.  The density of partons is higher in central
collisions at midrapidity and increases with energy, hence the baryon
to meson ratio would be highest in central collisions at midrapidity,
and increase with energy, - thereby explaining the characteristics of
the enhancement as seen in this conference.

\begin{figure}
\centering
\begin{tabular}{cc}
\epsfig{file=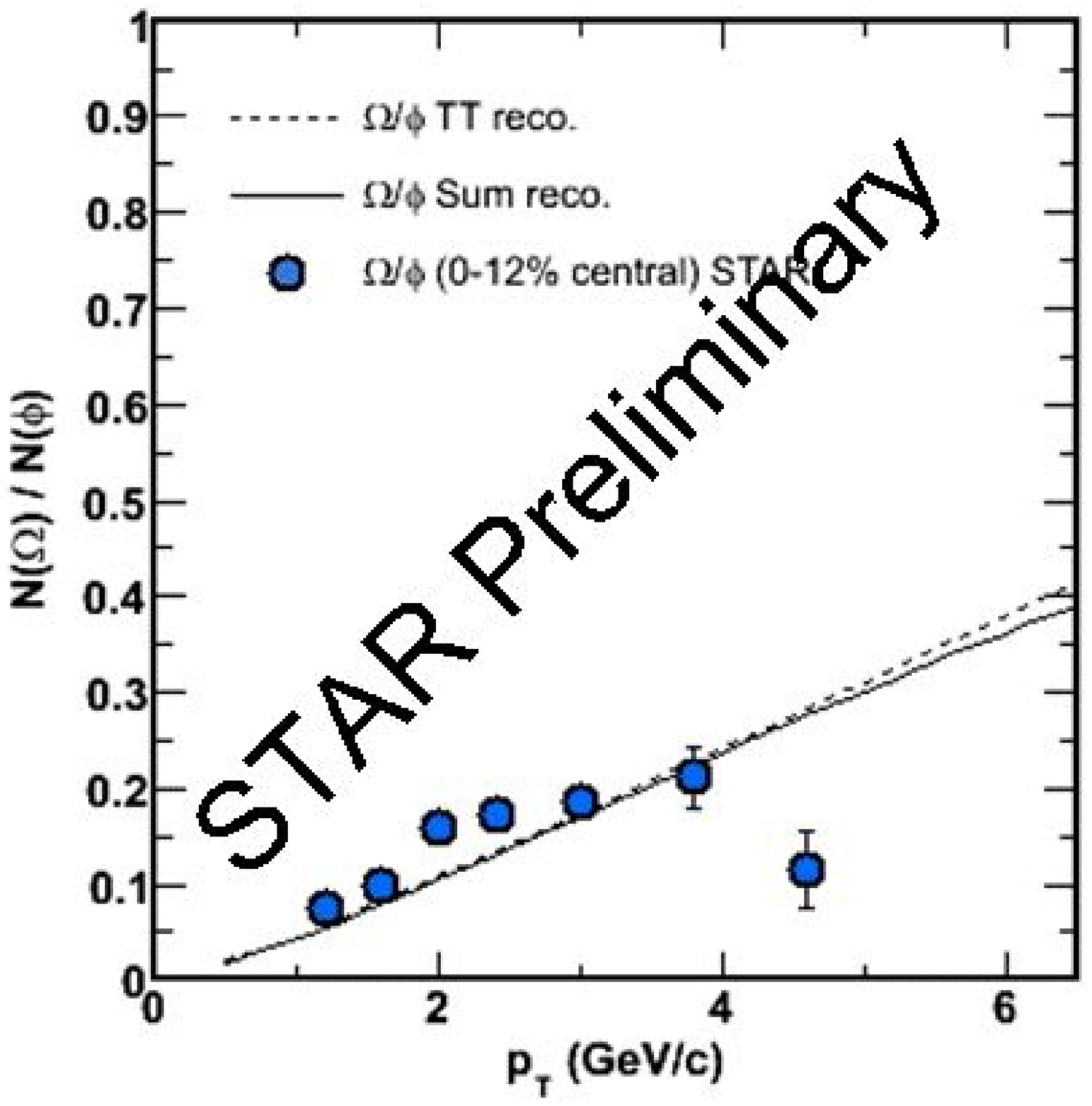,width=3in,clip=} & 
\epsfig{file=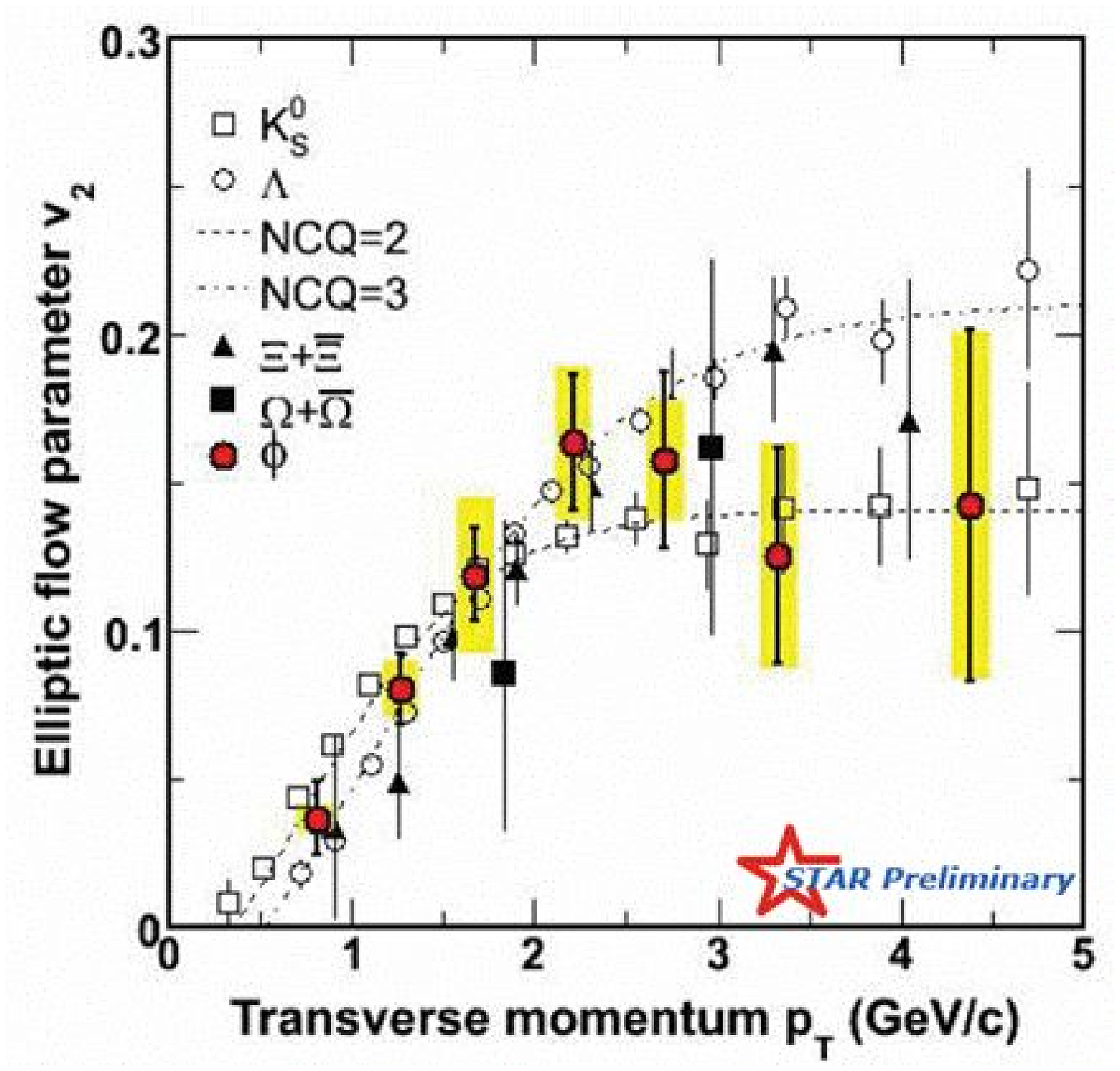,width=3in,clip=} 
\\
(a) & (b) \\
\end{tabular}
  \caption{ (a) Ratio of the yields, $\Omega/\phi$ as a function
of transverse momentum. (b) The elliptic flow, v$_2$ for the $\phi$ meson
as compared to other baryons and mesons. Note that the $\phi$ follows
the other mesons, despite the fact that its mass is more like that of
a proton or $\Lambda$}.
  \label{fig:stphi}
\end{figure}

The simplest of the recombination models assumes that partons which
recombine all come from a thermal distribution. This assumption,
however is contradicted by data from the PHENIX experiment, where
``jetlike'' correlations of baryons in the p$_T$ range of 2-5 GeV/c
are seen {\cite{pidjet}}. An alternate model of of recombination has
been proposed, by Hwa and collaborators {\cite{recomb}} in which essentially all hadronization is
explained by some form of recombination. In this picture fast partons
fragment into showers of partons and then recombine, either with other
partons within the shower with a result similar to standard
fragmentation functions, or with partons from the thermal bath. In
this model the $\Omega$ and the $\phi$ are dominantly composed of
``thermal'' partons as opposed to ``shower'' partons. This is true even out to p$_T$
$\sim$8 GeV. The model predicts that the ratio of $\Omega$ to $\phi$
yields should rise linearly with p$_T$ {\cite{hwaomegaphi}}. 
Fig.~\ref{fig:stphi}(a) shows the preliminary results presented by the
STAR collaboration for this ratio.  One can see that this linear
relationship holds, at least to about p$_T$$\sim$4 GeV/c. Whether the
final data point at 4.5 GeV/c is an indication of a breakdown of the
model, or a problem with the data is yet to be seen.
  
\section{Elliptic Flow, Viscosity and Recombination}

A third major surprise from RHIC was the large elliptic flow. 
The elliptic flow is characterized by the second
Fourier coefficient of the azimuthal anisotropy of the momentum spectra, 
v$_2$. 
Elliptic flow is generated in non-central collisions from the spatial 
anisotropy of the system. Pressure gradients
convert this spatial anisotropy into momentum anisotropy. 
Elliptic flow gives us a great deal of information on bulk properties, 
in particular it can give us a measure of the initial time
of thermalization and the viscosity of the system. In addition the analysis
of the elliptic flow of identified particles give additional
support for hadronization via recombination.

Let us consider  the time of thermalization.  If
the time of initial thermalization is late and the system free streams
for a period of time, any spatial anisotropy is lost and elliptic flow
is not developed. However, rapid thermalization will develop early
pressure gradients when the spatial anisotropy is at a maximum, which
then get converted intro large values of v$_2$. Detailed hydrodynamic
calculations assuming zero viscosity compared to RHIC results give
thermalization time of between 0.6 and 1 fm/c (i.e. between regions 2
and 3 in Fig.~\ref{fig:qgphistory}) and initial energy densities of
between 15 and 25 GeV/fm$^3$ {\cite{whitepapers}} consistent with the
energy density calculated in the parton energy loss calculations.

A great many results on flow were presented at this conference
on a variety of particles. There are three basic regions of elliptic flow.
\begin{enumerate}
\item the high p$_T$ region above 5-6 GeV where v$_2$ is generated by energy
loss of fast partons passing though a spatially anisotropic region. Note
that this is really not a ``flow'' in the sense of it being a collective
movement of particles at all, but is simply the result of the almond shape
of the initial state coupled with parton energy loss through the medium.
\item the region between 2 and 5 GeV/c transverse momentum where the elliptic
flow is generated in the partonic phase, ie. regions 3 and 4 of 
Fig.~\ref{fig:qgphistory}. For the study of the bulk properties
of the sQGP and recombination, this is the most interesting region.
\item the low momentum region below 2 GeV, where we will assume elliptic
flow is substantially affected
by rescattering in the hadronic phase (region 4 and 5) The importance of 
the hadronic phase 
to the generation of elliptic flow is unclear. The question of the amount
of spatial anisotropy remaining in the system at a time of several fm/c 
is one that must be answered quantitatively.
\end{enumerate}

One of the strongest pieces of evidence in favor of a recombination model
is that the elliptic flow of baryons and mesons can be rescaled
by the number of constituent quarks and shown to lie on an essentially
universal line{\cite{recomb}}. We will see this in a moment. A important
tests of 
this hypothesis is to examine the flow of the $\phi$ which
has a mass similar to the proton, but is a meson and hence is composed of two
constituent quarks. While the statistics is still low, it is clear from
Fig.~\ref{fig:stphi}(b) that the $\phi$ exhibits an elliptic flow
which is more like mesons than baryons and hence is dependent, not
on the mass, but on the number of constituent quarks. 

\begin{figure}
\centering
\begin{tabular}{cc}
\epsfig{file=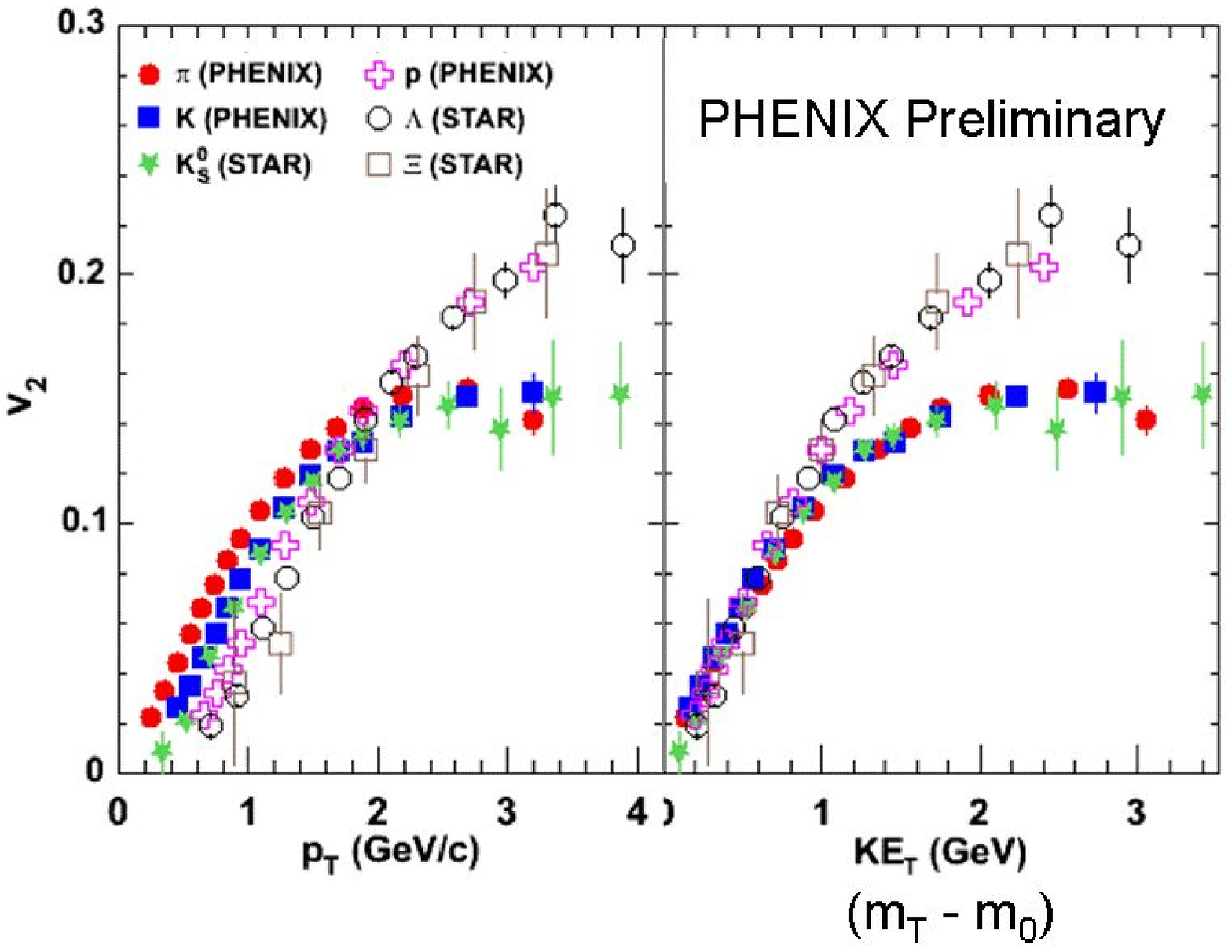,width=3in,clip=} & 
\epsfig{file=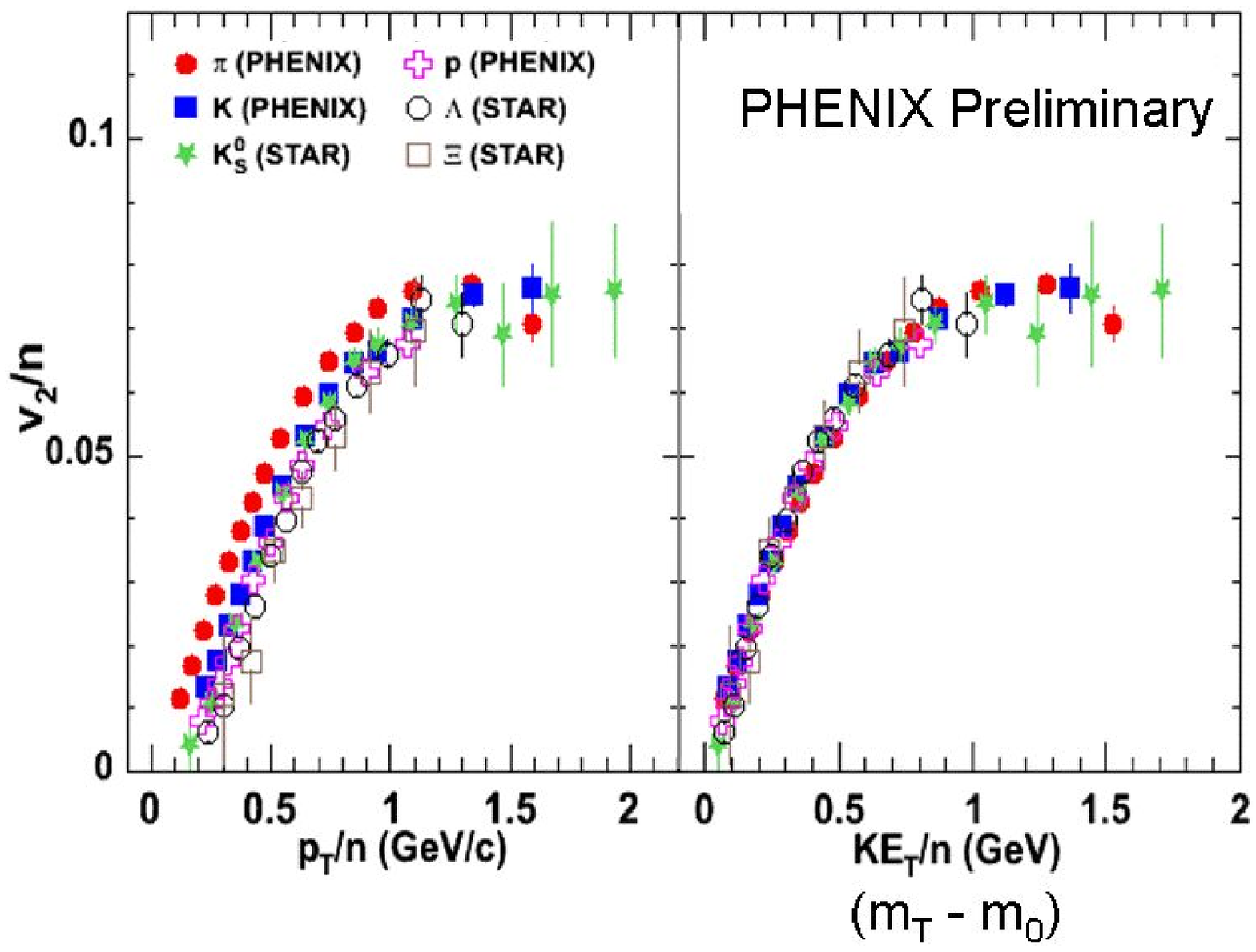,width=3in,clip=} \\
\ \ \ \ (a)\ \ \ \ \ \ \ \ \ \ \ \ \ \ \ \ \ \ (b) &\ \ \ \ (c)\ \ \ \ \ \ \ \ \ \ \ \ \ \ \ \ (d) \\
\end{tabular}
  \caption{v$_2$ for a variety of particles in minimum bias AuAu collisions
at $\sqrt{s}$=200 GeV  (a) as a function of p$_T$   (b) as a function of transverse kinetic energy, i.e. m$_T$-m$_0$ (c) v$_2$ scaled by the number of
constituent quarks as a function of p$_T$ (d) v$_2$ scaled by the number of
constituent quarks as a function of transverse kinetic energy.} 
  \label{fig:phflow}
\end{figure}

At this conference PHENIX has shown a way of presenting the data which 
gives some insight
into the mechanisms leading to elliptic flow - in particular the 
low and moderate transverse momentum
regions(Fig.~\ref{fig:phflow}). The argument is as follows. In the hadronic phase, one should plot
measurements as a function of the ``transverse kinetic energy'', i.e.
m$_T$-m$_0$. This may make some sense in that the pressure gradients 
in this phase are essentially due to the kinetic energy of the hadron
gas (assuming an ideal gas model which is certainly NOT correct in the partonic
phase). Fig.~\ref{fig:phflow} (b) shows that once this new variable is used
the low momentum values of the elliptic flow fall on a single line. 
The mass splitting seen in the data at low transverse momenta, are well 
described by hydrodynamic models {\cite{hydrofitv2}} which
impart an elliptic flow dependent on the mass of the particle.
If one further rescales
by the number of constituent quarks Fig.~\ref{fig:phflow} (d), one
sees the data collapse onto a single curve, giving support to the idea
that the elliptic flow, at least for momenta above about 2 GeV/c is 
developed in the partonic phase. 

\begin{figure}
  \begin{center}
    \leavevmode
    \epsfig{file=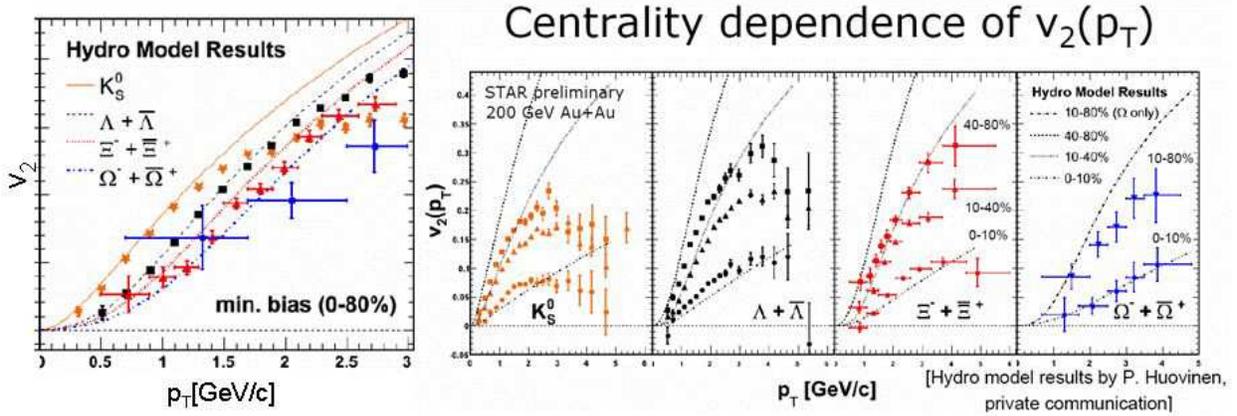, width=6.5in}
  \end{center}
  \caption{(left) v$_2$ for minimum bias collisions in the low 
p$_T$ region for a variety of particles in $\sqrt{s}$=200 GeV AuAu collisions
as compared to a hydrodynamical calculation. Note that the model reproduces
the mass splitting below about 1.5 GeV/c. (right) v$_2$ for various centralities as compared to a hydrodynamical model. }
  \label{fig:stoldenburg1}
\end{figure}

\begin{figure}
  \begin{center}
    \leavevmode
    \epsfig{file=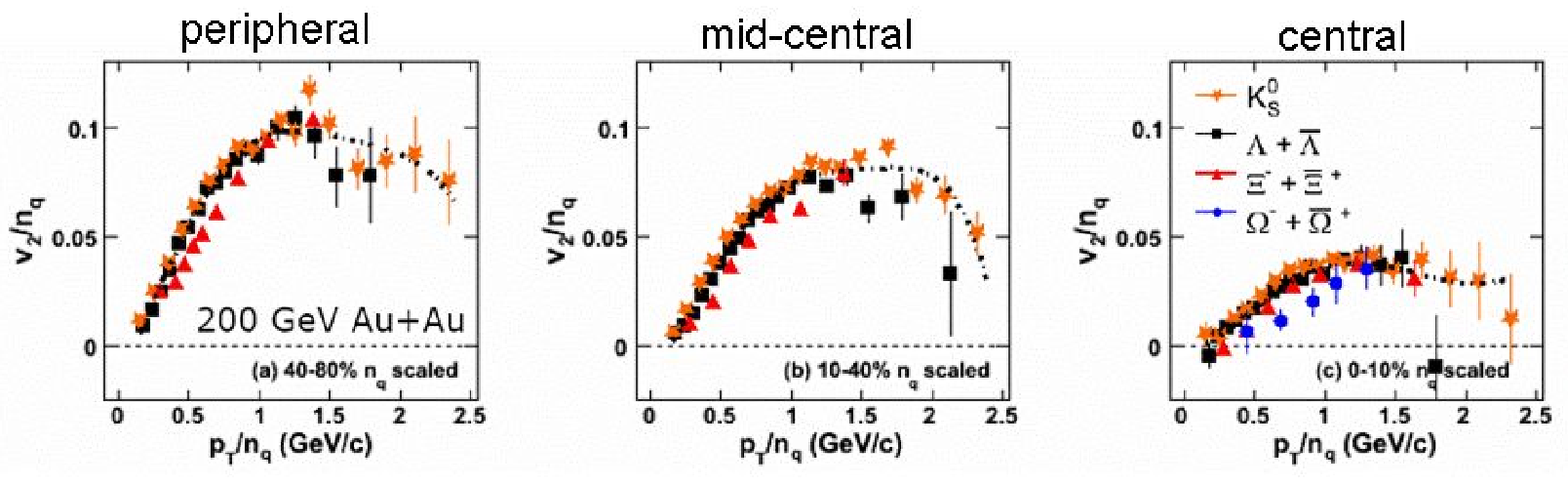, width=6.5in}
  \end{center}
  \caption{v$_2$/n for various centralities. Preliminary results from STAR. }
  \label{fig:stoldenburg2}
\end{figure}

The data on elliptic flow shown thus far, has been from minimum bias events. 
A more detailed comparison between hydrodynamical models and centrality 
selected data shows rather poor agreement with the data, however,
quark constituent number scaling continues to work 
(see Fig.~\ref{fig:stoldenburg1} and Fig.~\ref{fig:stoldenburg2}){\cite{oldenburg}}.

The disagreement of the hydrodynamic models with the data
 may simply indicate that the
initial conditions taken in those models must be adjusted, or it may be
that such comparisons pose a problem for the theory - at least
the portion of the hydrodynamics models in which the medium is
converted to massive hadrons - 
the so called ``Cooper-Frye'' mechanism{\cite{hydrocooperfrye}}.

\begin{figure}
  \begin{center}
    \leavevmode
    \epsfig{file=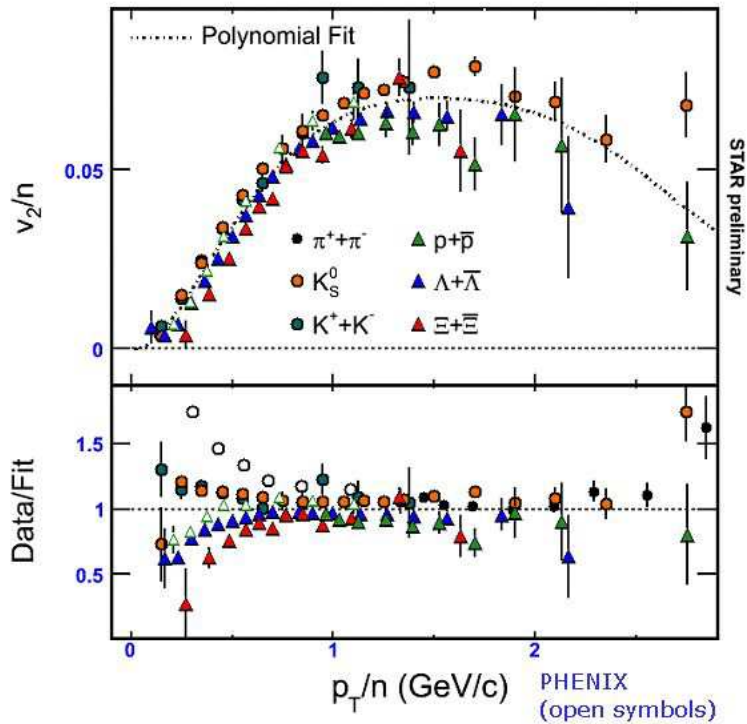, width=4in}
  \end{center}
  \caption{v$_2$/n for 200 GeV minimum bias AuAu events, with
detail, showing that mesons tend to lie above the fit line
and baryons below.}
  \label{fig:sthigherfock}
\end{figure}

The recombination model works remarkably well,however(Fig.~\ref{fig:stoldenburg2}). The question
arises - exactly what are the ``constituent quarks''? What has
happened to the gluons?  Muller et al, have attempted to make headway
on this question, by improving their model{\cite{recomb2}}. Quark
wave functions are augmented by adding corrections for gluons. This
breaks the perfect quark number scaling with mesons having a slightly
higher value for v$_2$/n, where n is the number of constituent quarks
in the hadron. The STAR collaboration has examined the data carefully
(Fig.~\ref{fig:sthigherfock}) and indeed the mesons fall slightly
above the baryons in the region of p$_T$/n 1-3 GeV where the model is
valid.

One of the tasks that lies ahead is a deeper and more quantitative
understanding of the degrees of the freedom of the sQGP. A major
theoretical step has been taken towards this end, using correlation
functions between baryon number, charge and strangeness
{\cite{majumdergavai}} in examining lattice results. They conclude
that the degrees of freedom on the lattice above the phase transition
have the quantum numbers of quarks - ``quasi-quarks'', which of course
reinforces the previous conclusions with respect to recombination. 
This theoretical result, together
with the data are perhaps the most important conclusion of the
conference. We are now beginning to understand the degrees of freedom
of the sQGP - they seem to be ``quasi-quarks'', which can be viewed as
quarks - dressed with gluons and other higher ``Fock states'' - e.g.
quark-anti-quark pairs. Why these are the degrees of freedom is still
to be understood.

\subsection{Elliptic Flow of Heavy Quarks}

In crude terms, viscosity is a measure of the tendency of momentum to
dissipate in a system. Large cross sections, and small mean free paths
are characteristic of small viscosity. In such systems, spatial
anisotropies which give rise to pressure gradients will be efficiently
converted into momentum anisotropy, thereby resulting in a large
elliptic flow.  The fact that the elliptic flow is so large -
essentially being described by models with zero viscosity - means that
the system is strongly interacting; hence the name sQGP. We would like
to quantify these statements however by comparing measurements of the
strength of the self interaction and the value of the velocity with
theory. The initial step is the obvious one - that is to figuratively
throw a stone into the strongly interacting fluid and see if it
moves. The stone we use is heavy quarks - charm and bottom.

\begin{figure}
  \begin{center}
    \leavevmode
    \epsfig{file=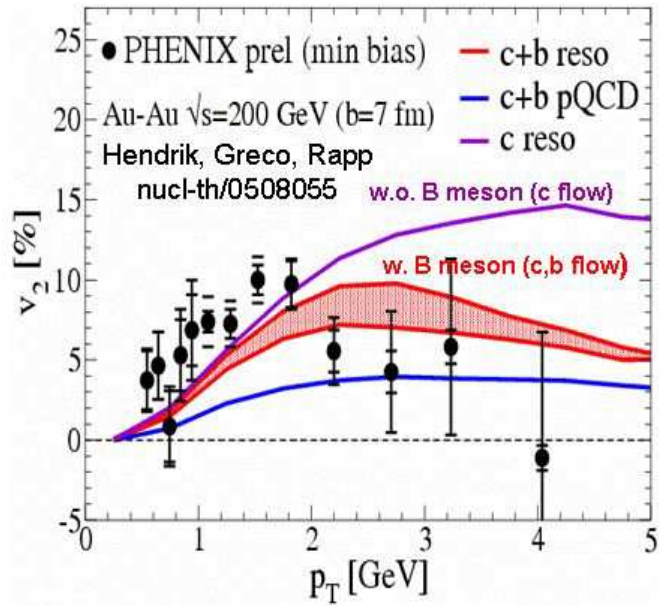, width=3.5in}
  \end{center}
  \caption{v$_2$ of non-photonic electrons as compared to a model. 
The shaded fit assumes that the source of electrons is both charm and
bottom and that both are interacting with the medium and flow. The flow
of bottom quarks is considerably smaller than that of charm quarks because
of the mass.}
  \label{fig:phcharmflow}
\end{figure}

A first measurement of the elliptic flow of heavy quarks has been made
by the PHENIX collaboration by examining v$_2$ of ``non-photonic''
electrons. Fig.~\ref{fig:phcharmflow} shows the data as compared to
a simple model where heavy quarks were assumed be the source of
electrons. It is clear that some amount of charm flow is necessary to
explain the data.  In the model, light quarks were assumed to flow
with a magnitude consistent with the previous data on hadronic 
elliptic flow.  A mechanism (resonant
scattering) for introducing strong interactions between the light
quarks and heavy quarks is invoked where these interactions are
strong enough to reduce heavy quark thermalization times by a factor
of 3 as compared to pQCD{\cite{rapp}}. The flow of bottom quarks is of
course considerably less than that of charm quarks due to their mass.
The various lines are for a variety of cases as explained in the
caption.  We can conclude from this comparison, that charm flows,
and that the interactions
between constituents are indeed strong and the viscosity, small.

\section{The Colored Glass Condensate}
\begin{figure}
\centering
\begin{tabular}{cc}
\epsfig{file=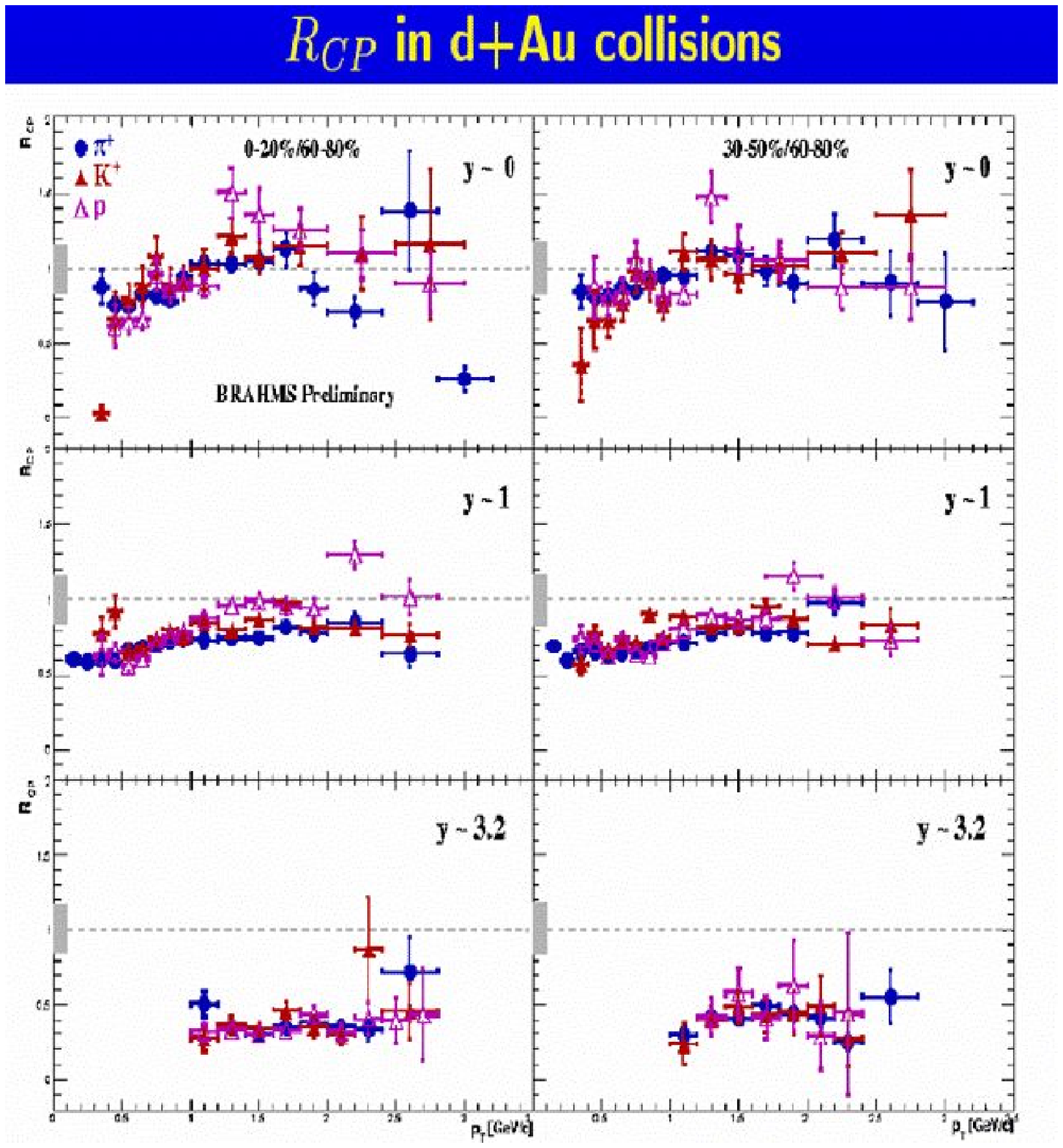,width=3in,clip=} & 
\epsfig{file=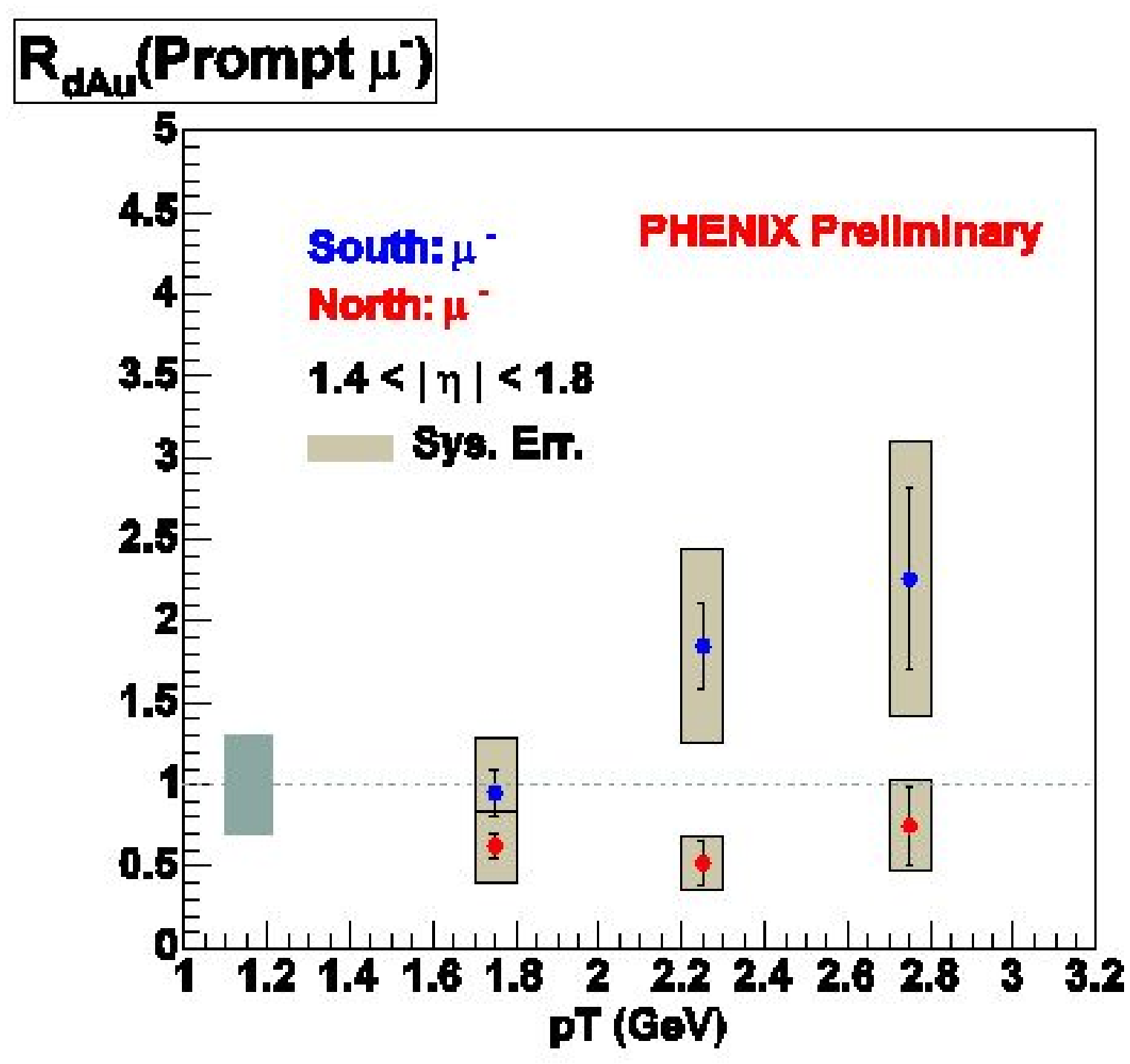,width=3in,clip=} \\
(a) & (b)\\
\end{tabular}
  \caption{(a) R$_{CP}$ for pions, kaons and protons in deuteron-gold collisions
at $\sqrt{s}$=200 GeV from the BRAHMS collaboration. R$_{CP}$ is similar
to R$_{AA}$ with peripheral collisions replacing pp in the ratio. The left column is a ratio of 0-30\% centrality divided by 60-80\%, the right shows 30-50\% centrality divided by 60-80\%. Higher values of centralit correspond to more 
peripheral collisions. (b) R$_{dA}$ for prompt muons in $\sqrt{s}$=200 GeV
d+Au collisions.  R$_{dA}$ is similar to R$_{AA}$ but for deuteron-nucleus collisions. The north arm is in the deuteron direction (lower data points).  }
  \label{fig:cgc}
\end{figure}

One of the most attractive models of the initial conditions for the
formation of the sQGP, is a saturated gluon state, or the Colored
Glass Condensate {\cite{cgc}}. Proton-nucleus or deuteron-nucleus
collisions provide a laboratory to isolate these effects from the
sQGP. A suppression of the transverse momentum spectra is predicted.
(It is important to note that in AuAu collisions at central rapidity,
such effects are not the cause of the suppression of high p$_T$
particles. High momentum hadrons showed no suppression in deuteron
nucleus collisions{\cite{dAu}}. Midrapidity collisions do not probe 
the relevant regions of Bjorken-x for saturation to be significant.
Saturation at RHIC must be studied at forward rapidities.)  
The suppression should be stronger
for higher rapidities and more central collisions.  Since the effect
is primarily concerned with gluons, the suppression should be independent
of particle type. Fig.~\ref{fig:cgc} (a) shows data from the BRAHMS
collaboration showing the suppression as a function of centrality for
a variety of rapidities for pions, kaons and protons, and as
predicted, the effect is independent of particle type and the
suppression increases with centrality and rapidity.  PHENIX has
measured prompt muons, primarily from heavy quarks which show a
similar suppression(Fig.~\ref{fig:cgc} (b)) As has been seen
previously for light hadrons, the PHENIX collaboration also observes a
rather strong enhancement in the Au direction.

\section{Deconfinement and Screening}
Thus far, the data fits remarkably well into the general picture
presented at the beginning. I have sidestepped the actual experimental
verification of the phase transition itself - either deconfinement or
chiral restoration.  The traditional signature for deconfinement -
J/$\psi$ suppression has left us with a puzzle. There are serious
doubts as to whether it can tell us anything about deconfinement. The
problems are both theoretical and experimental. On the theory side,
recent lattice calculations indicate that the J/$\psi$ remains bound
until well above the critical temperature. One might then hope to use
other charmonium states which ``melt'' at lower temperatures, such as
the $\psi$' or the $\chi_C$. However there is a disagreement in the
theoretical community as to the mechanism for the ``melting'' of the
J/$\psi$. Traditionally Debye screening was thought to be the primary
mechanism. The fact that the sQGP is strongly interacting has already
taught us that screening lengths are much longer than had been
originally thought - a short screening length would have resulted in a
quasi-free QGP. Breakup of the J/$\psi$ from the collision from hard
gluons should also lead to the suppression of the J$\psi$. There is
also theoretical disagreement about whether a the interaction between
partons - in particular for charmonium - can be modeled by a
potential. Finally, given the success of the recombination picture,
there is a real possibility that charmonium states can be regenerated.
While theoretical uncertainties make experimentalists somewhat
uncomfortable, such upheaval is not surprising, given that the
paradigm for the QGP has undergone such change in the past 2 years.

\begin{figure}
\centering
\begin{tabular}{cc}
\epsfig{file=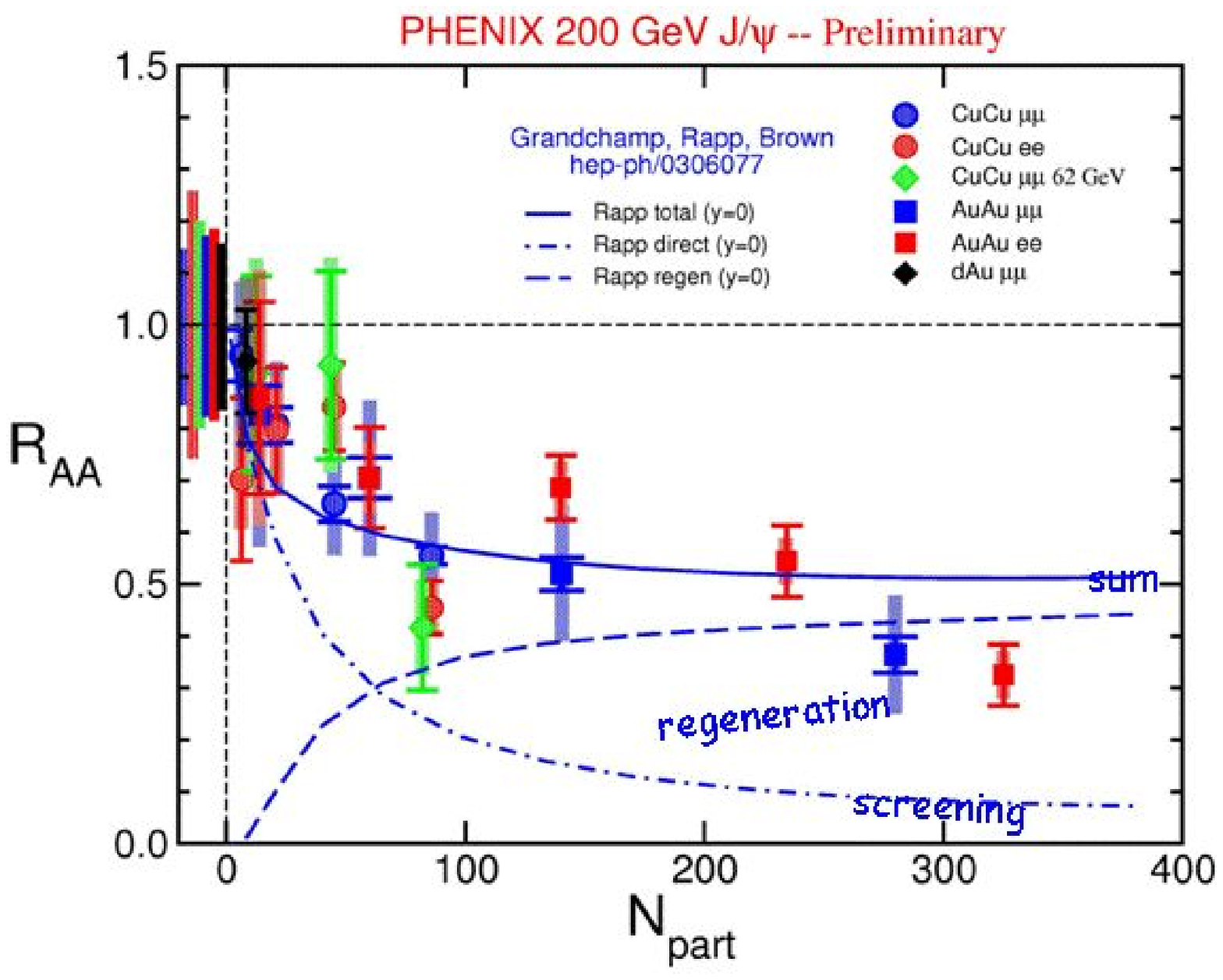,width=3.3in,clip=} & 
\epsfig{file=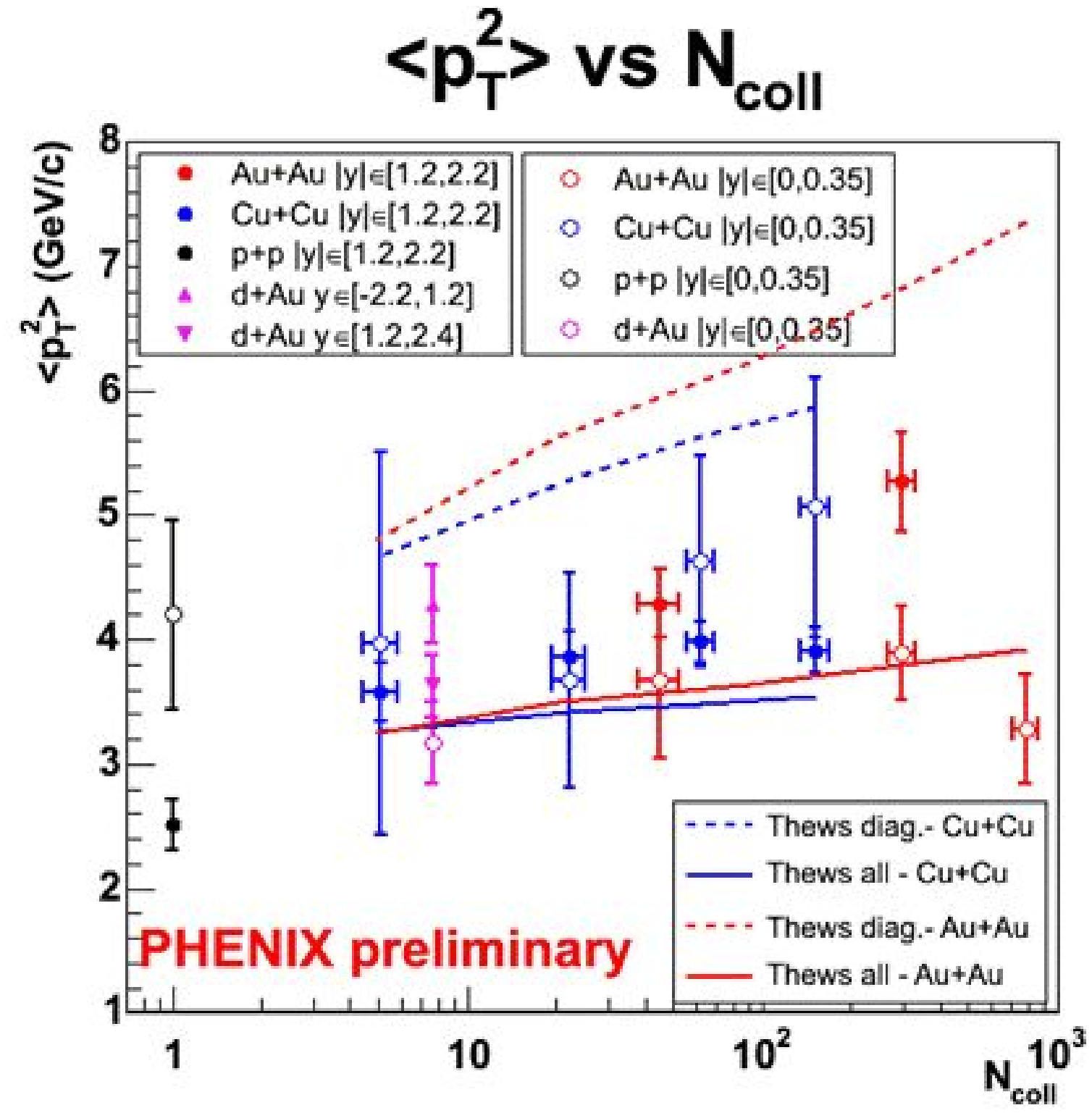,width=2.7in,clip=} \\
(a) & (b)\\
\end{tabular}
  \caption{(a) R$_{AA}$ for the J$\psi$ as compared to model including
screening and regeneration.   (b) $\langle p_T^2\rangle$ for the J/$psi$ as 
as compared to a model. The dashed lines is without recombination effects,
the solid lines include recombination.}
  \label{fig:phjpsi}
\end{figure}

\begin{figure}
\centering
\begin{tabular}{cc}
\epsfig{file=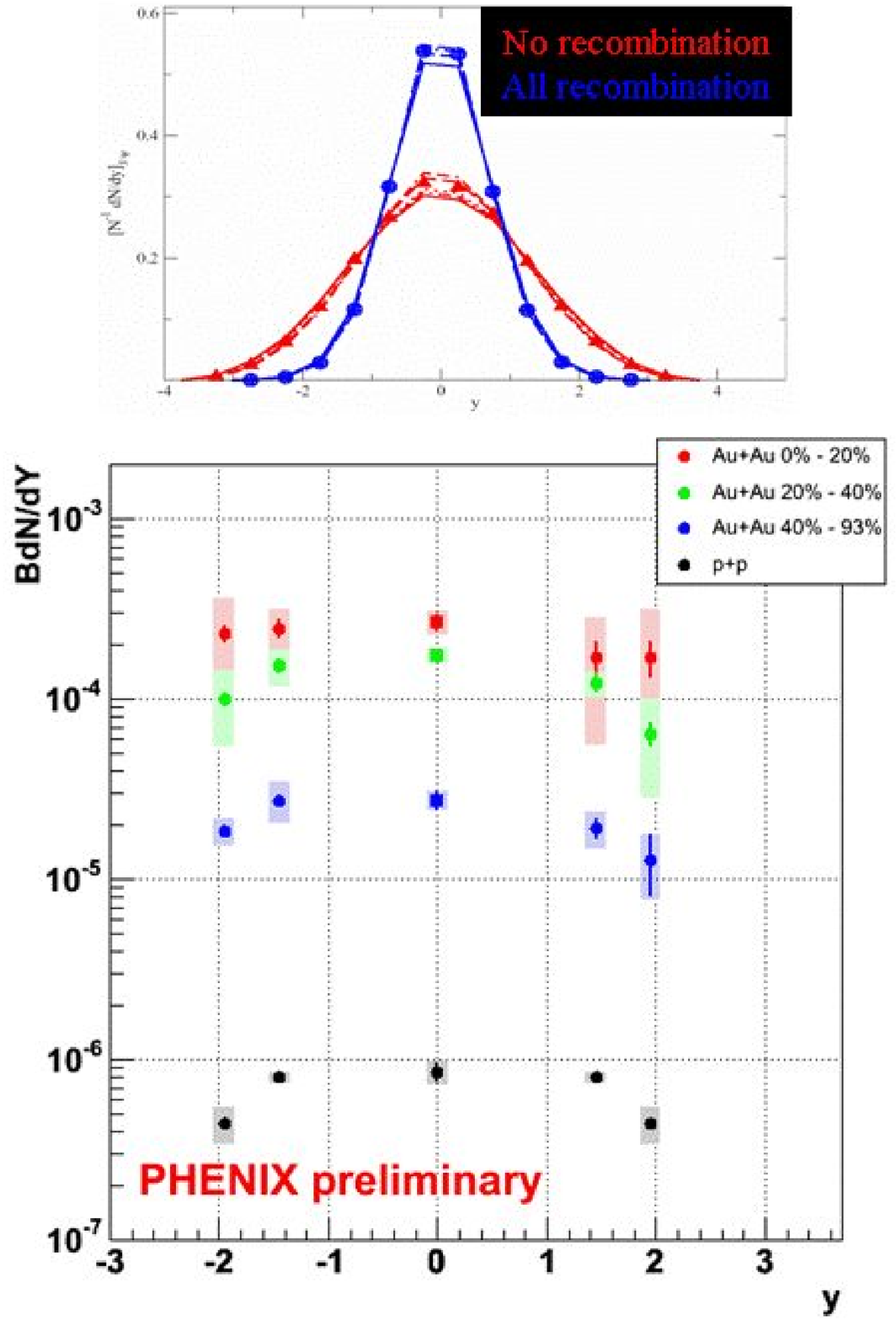,width=3in,clip=} & 
\epsfig{file=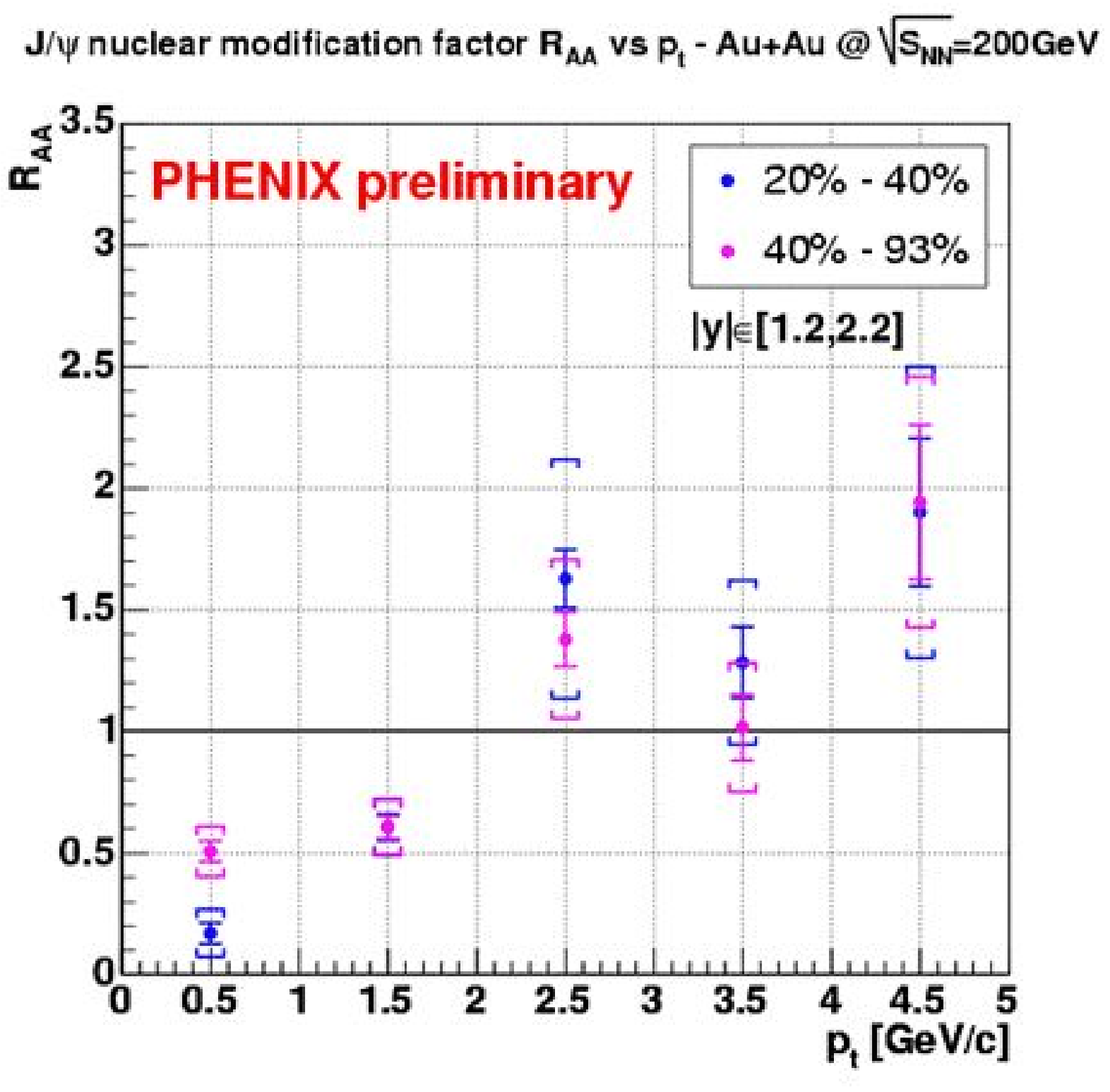,width=3in,clip=} \\
(a) & (b)\\
\end{tabular}
 \caption{(a) The top plot shows the effect of recombination in narrowing
the rapidity distribution for J/$\psi$'s in heavy ion collisions as compared
to pp collisions{\cite{thews}}
. The bottom plot is data from PHENIX showing
that there is no narrowing of the rapidity distribution in going
from pp to central AuAu collisions.  (b) R$_{AA}$ for the J/$\psi$ 
as a function of p$_T$ showing that the suppression is at low transverse momentum.}
  \label{fig:phjpsiypt}
\end{figure}

Given the state of our understanding,
 the data on the J/$\psi$ provide us with more
questions than answers. The suppression seen by the PHENIX
collaboration is shown in Fig.~\ref{fig:phjpsi}(a). This is similar in
magnitude to the effect seen in the SPS. A prediction from the
screening mechanism underpredicts the data considerably. Models which
include regeneration, however, do a reasonable job of reproducing the
data.  One can then look at the average of transverse momentum
squared.  If the J/$\psi$ were a result of recombination, this would
tend to reduce the transverse momentum as seen from the models in
Fig.~\ref{fig:phjpsi}(b). As one can see, the data favors
recombination.  However, such models would also predict that the
rapidity distribution to narrow. The data show no such effect(Fig.~\ref{fig:phjpsiypt}(a)). Lastly (Fig.~\ref{fig:phjpsiypt}(b))
shows R$_{AA}$ for the J/$\psi$ as a function of p$_T$ showing
that the suppression occurs at low p$_T$. This would be consistent
with a picture in which slower particles are suppressed (screened)
 in the medium and fast particles escape. While no firm conclusions
can be made about deconfinement at the moment - it is imperative
for both theorists and experimentalists to pursue these measurements.
Puzzling and contradictory results often lead to new ideas. 

\section{Conclusions}
What are the new things we have learned from the data presented at this conference? 
\begin{itemize}
\item Heavy quark energy loss is now well established, as both STAR
and PHENIX have observed the suppression of high momentum ``non-photonic''
electrons. Critical measurements are left to be made in which
the contribution from charm and bottom quarks can be separated.
\item Charm quarks flow, giving us further evidence that the viscosity
is small and that interactions are large.
\item Recombination continues to find experimental support. All particles
seem to obey constituent quark scaling in both the baryon to meson ratios
as well as the elliptic flow. In addition STAR has showed us some evidence
for ``fine structure'' in the v$_2$/n plots which are consistent with
theoretical ideas of dressed ``quasi quarks'' being the degrees of freedom
of the sQGP. 
\item PHENIX has shown the suppression of heavy quarks in dAu collisions
at forward rapidity as predicted by CGC models.
\end{itemize}

The field of relativistic heavy ions is blooming. We have begun
to understand some things, but there is a great deal left to do. Most of our
conclusions remain rather qualitative in nature - we would
like more quantitative results. In order for this to happen both theorists
and experimentalist must work together on common questions. For instance -
we know now that this viscosity is small - but exactly \it{how small}?? 
\rm With upgraded detectors and an expected increase in luminosity for RHIC,
experimentalists will need to measure the elliptic flow of identified
charm and bottom mesons. Theorists will need to be able to associate
these results with a numerical viscosity of the medium. 

There are broader questions as well. The phase transition itself remains
elusive - i.e. we would like to see direct evidence of deconfinement
and/or chiral symmetry restoration, measure the transition
temperature, and find the critical point. 

In closing - I would like to reiterate one major advance I think we have
seen in this conference - and that is, that we are beginning to get
some idea of what the degrees of freedom are of the sQGP. All experimental
evidence seems to point to ``quasi-quarks'' (for lack of a better name).
Lattice measurements of susceptibilities seem to confirm this.
We must seek to understand these degrees of freedom - in particular-
we would like to know whether these degrees of freedom somehow are
connected to the fact that the
stuff we have created at RHIC is so strongly interacting.


\end{document}